\documentclass[onecolumn,floatfix,showpacs,nofootinbib,11pt]{revtex4}
\usepackage{amssymb,amsmath,dsfont,graphicx}
\usepackage{color}
\usepackage[colorlinks,hyperindex]{hyperref}

\definecolor{green1}{RGB}{0,128,0} 
\hypersetup{hidelinks,backref=true,pagebackref=true,hyperindex=true,colorlinks=true,breaklinks=true,urlcolor= blue}
\hypersetup{%
  colorlinks = true,
  linkcolor  = blue,
  citecolor = green1,
}
\usepackage{bookmark}
\usepackage{hyperref,color,xcolor}
\hypersetup{hidelinks,hyperindex=true,colorlinks=true,breaklinks=true,urlcolor= blue}
\hypersetup{%
  colorlinks = true,
  linkcolor  = blue
}

\usepackage{makeidx}
\newcommand{\beq}{\begin{eqnarray}}\newcommand{\benu}{\begin{enumerate}}\newcommand{\enu}{\end{enumerate}}
\newcommand{\eeq}{\end{eqnarray}}

\newcommand{\el}{^}
\newcommand{\lp}{\ell_{\rm p}}
\newcommand{\mpl}{m_{\rm p}}

\newcommand{\ex}{\text{e}}
\DeclareMathOperator{\x}{x}
\newcommand{\be}{\begin{equation}}
\newcommand{\ee}{\end{equation}}

\newcommand{\ba}{\begin{eqnarray}}
\newcommand{\ea}{\end{eqnarray}}

\begin{document}
\title{Monopoles on  string-like models and the Coulomb's law}
\author{D. M. Dantas}
\email{davi@fisica.ufc.br}
\affiliation{Departamento de F\'{i}sica - Universidade Federal do Cear\'{a}  C.P. 6030, 60455-760, 
Fortaleza, Cear\'{a}, Brazil.}
\author{R.~da~Rocha}
\email{roldao.rocha@ufabc.edu.br}
\affiliation{Centro de Matem\'atica, Computa\c c\~ao e Cogni\c c\~ao, Universidade Federal do ABC, 09210-580,
Santo Andr\'e, Brazil.}
\pacs{11.25.-w, 03.50.-z, 14.80.Hv}

\author{C. A. S. Almeida}
\email{carlos@fisica.ufc.br}
\affiliation{Departamento de F\'{i}sica - Universidade Federal do Cear\'{a}  C.P. 6030, 60455-760, 
Fortaleza, Cear\'{a}, Brazil.}

\begin{abstract}
{The t'Hooft-Polyakov monopole mass can be substantially altered, in the thick GS and HC brane-world setup, and can be employed to constrain the brane thickness limit. In this work, we comprise a brief review regarding gauge fields localization in the string-like six dimensional brane-world models setup. The correction to the Coulomb's law in two models is studied.  Besides, the monopole features are investigated from the point of view of the gauge fields localization in the string-like brane-worlds and its thickness.}
\end{abstract}
\maketitle

\section{Introduction}

The main aim of this work is to deal with the mass of the monopole, in string-like models on six dimensions. It is remarkable to mention that in the scope of applications of brane-worlds models,   several recent works perform the study of the informational entropy \cite{ce1,ce2,ce3,ce4, Braga:2016wzx, Bernardini:2016qit, Bernardini:2016hvx} underlying these scenarios, using Einstein-dilaton systems to derive phenomenologically compatible data. These models are used to respond to LHCb anomalies \cite{lhcb1, lhcb2}, and several others applications, in  the context of black holes \cite{black1,black2,black3}, dark matter \cite{dark1} and inflationary universe theories \cite{inf1, inf2, inf3}.

Six-dimensional (6D) anti-de Sitter ($AdS_6$) brane-worlds \cite{GS1, Silva:2012yj} have some interesting features, as the remarkable advantage of trapping the massless gauge fields with only gravity \cite{Oda1,Costa:2015dva, Diego, DM3, Coni2}, in opposition to what occurs in 5D domain wall models \cite{bajc} and DPG models \cite{Hinterbichler:2010xn}, and minimally coupled Dirac fermions  \cite{Liu1, DM3, ferm1,ferm2, elko6DEPL}, even when  thin brane models are considered.  In the context of Lorentz violation scenarios in brane-worlds, the  massless four-dimensional (4D) graviton can be only confined  in 6D models \cite{victor6d}. Fermions fields in 6D have also prominent applications to  the generation of fundamental fermions and neutrinos \cite{Merab, Frere}.

Moreover, the thickness of the brane plays an important role. Smooth brane-world have been investigated, presenting a thickness $\Delta$ that drive deviations from the 4D Newton's law
 on scales of such a magnitude \cite{Bazeia:2003aw,Barbosa-Cendejas:2013cxa, DM6}. Current   torsion-balance experiments yield the upper constraint 
$\Delta \lesssim 44\,\mu$m~\cite{kapp} for the brane thickness, that also have to be as big as the 5D Planck scale, 
$\Delta\gtrapprox 2 \times 10^{-21}\,$cm.
The 4D Planck mass  $\mpl\simeq 2.2\times 10^{-8}\,$kg and the length prescribing the
4D Newton's constant $G_{\rm N}=\lp/\mpl$ are the main ingredients in our analysis, and throughout this paper, natural units shall be employed.
 Although widely investigated in 5D models, 6D models lack still such kind of phenomenological approach. Notwithstanding, in what follows, 
the 6th dimension does not play any additional role on fixing the 5D thickness in 6D models, being our approach here equivalent to the 5D ones in the literature.

't Hooft-Polyakov monopoles are physical solutions 
corresponding to localized topological solitons with finite energy,  firstly suggested in the context of Georgi-Glashow models \cite{glas}. SU(2) monopoles are allowed to place on a warped thick brane-world. The usual monopole radius was shown to be affected by the warped geometry with sine-Gordon potentials ruling the scalar field that generates the thick brane, in Ref. \cite{epl2012}. As observed in that reference, the more distant from the thick brane core the monopole, 
 the bigger its radius. Hence, one can assert that the monopole radius decreases as a function of the thick brane evanescence out of the 4th dimension into 5D, exactly attaining a  maximum value at the thick brane core. In fact, the naked monopole radius, i. e., without the warp multiplicative factor, is dressed by the spontaneous symmetry breaking parameter in a Higgs-like potential, in a thick brane setup. This radius is inversely proportional to the monopole mass \cite{RSI}. As the dressed --  visible  -- monopole radius does depend on the monopole position, an upper limit on the thick brane width can be hence established. 
 
 Effective monopoles were considered in Ref. \cite{epl2012}, without  effects of the gravitational back-reaction, 
 those which have survived out the period of inflation. Therefore, the brane-world cosmology is here  assumed not to be altered with respect to the usual cosmological scenario. More precisely, the current existence of $SU(2)$ monopoles is also proposed in Ref. \cite{Matsuda:2004nx}, where a monopole can be produced as a $D_3$ brane, whose  tension was shown to coincide with the mass of a monopole in the
effective action. The brane inflation may promote the formation of monopoles after the brane inflation, which are not negligible  in models of brane cosmology \cite{Matsuda:2004nx}.

In the present paper, we analyse the effect of the monopole mass on the thickness of two string-like models. The Gherghetta--Shaposhnikov (GS)  \cite{GS1} model and the Hamilton Cigar (HC) \cite{Silva:2012yj} model. Additionally, the study of thickness parameter by the configurational entropy for both GS and HC model was already performed in the Ref. \cite{ce4}.
This paper is structured as follows: in Sect.  \ref{sec-gauge},  we present a brief review of string-like models in 6D, their prominent features concerning the gauge vector fields localization and the correction to the Coulomb's law as well. In Sect. \ref{sec-mono}, we use the fact that the monopoles have not been observed, in order to delimit the brane thickness. The main concepts and results present in this paper are summarized in the concluding section \ref{sec-conclusions}.

\section{The six-dimensional string-like models and the vector gauge localization}\label{sec-gauge}
\subsection{Review of gauge fields on string-like models}
The localization of  vector gauge fields in six-dimensional (6D) scenarios has been  already  performed in several works \cite{Oda1,Costa:2015dva, Diego, DM3, Coni2}. On the String-like models, the confinement of vector gauge fields is performed via gravitational coupling,  without any additional scalar coupling \cite{Oda1}. Thus, the vector field can be localized on the string-like defect just with the exponentially decreasing warp factor, in opposition to what occurs in 5D models \cite{bajc, Hinterbichler:2010xn}. In this section, we briefly review  the localization of the $U(1)$ gauge fields in these string-like models.

The Randall--Sundrum-like metric ansatz for these 6D string-like models can be described by \cite{GS1, Oda1, Silva:2012yj}
\begin{gather}
ds^2_6=\sigma(r)\eta_{\mu \nu} dx^{\mu} dx^{\nu}+ dr^2+\gamma(r) d\theta^2 ,
\label{6dmetric}
\end{gather}
where the Minkowski signature is $\eta_{\mu \nu}= diag(-1,+1,+1,+1)$. The warp factors $\sigma(r)$ and $\gamma(r)=\beta(r)\sigma(r)$  exponentially decrease, with $\beta(r)$ being correlated to the compactification of the model \cite{GS1}. The coordinates range $r \in [0, \infty)$ and $\theta \in [0,2\pi)$.

In this work we analyse two $AdS_6$ models. The first one is the thin string-like model called Gergheta--Shaposhnikov model (GS) \cite{GS1}. The second one is the regular thick models called Hamilton Cigar (HC) \cite{Silva:2012yj}. The advantage of the GS model relies on the fact  that mostly of the results can be  analytically obtained. However, the regularity conditions  do not hold  in the GS model \cite{Silva:2012yj, Tyniakov}. We represent the warp factor of these models in the Table below, \ref{tab-wf}.
\begin{table}[h] %[!htb]
\begin{tabular}{|| l || c |c ||}
\hline\hline
   & GS  & HC\\
  \hline\hline
 $\sigma(r)$& \quad $\ex^{-cr}$ \quad& $\ex^{-cr+\tanh(cr)}$ \\
  \hline
  $\beta(r)$ & \quad $\mathfrak{R}_0^2$ \quad & \;\;$c^{-2}\tanh^2{(cr)}$ \\
  \hline\hline
\end{tabular}
\caption{Warp factors for the Gergheta--Shaposhnikov and the Hamilton Cigar models.}
\label{tab-wf}
\end{table}

 The thickness  parameter $c=\sqrt{-\frac{2}{5} \Lambda \kappa_6}>0$ is dependent of cosmological constant $\Lambda$ and also on the 6D Newton's constant $\kappa_6$ \cite{GS1}.  The $\mathfrak{R}_0$ is an arbitrary length scale parameter \cite{GS1}. Moreover, the $c$ thickness  parameter is  also correlated to the model Ricci  curvature,  being a pure $AdS_6$ space for the GS model $R=-\frac{15}{2}c^2$ \cite{GS1}. The brane is placed at the origin on the GS model, whereas it is slighted displaced from the origin  on the regular HC model \cite{Silva:2012yj}.  
 %${6.32564, {z -> 0.47944}}$

\begin{figure}[!htb] % Duas figuras lado a lado
%\begin{minipage}[t]{0.45 \linewidth}
        \centering
%    \begin{subfigure}[b]{0.50\textwidth}
                \includegraphics[width=.45\linewidth]{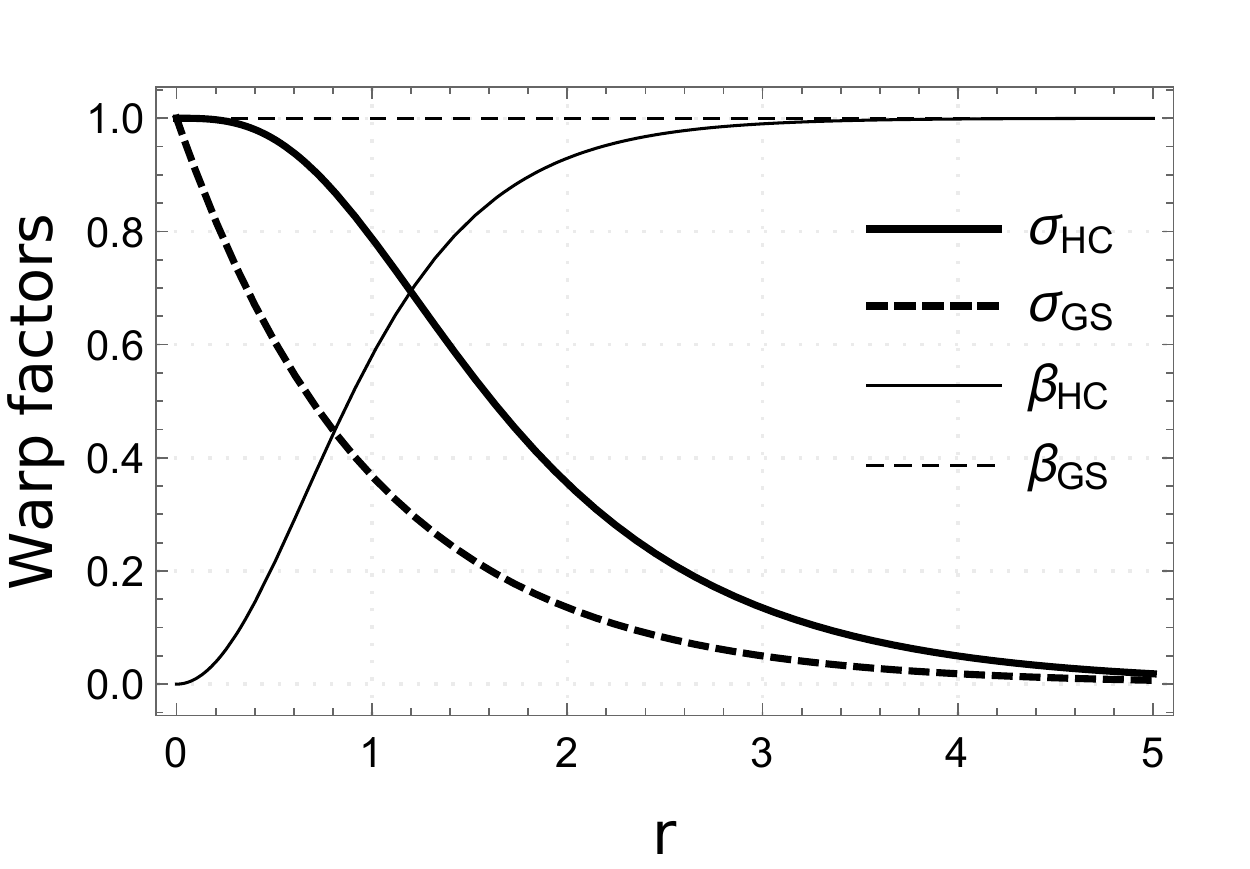}
                \caption{Warp factors for the HC (continuum lines) and GS (dashed lines) models. The $\sigma(r)$ are represented by thick lines, whereas the $\beta(r)$ are represented by thin lines. We use $c=1$ and $\mathfrak{R}_0=1$.}
\label{fig-warp}
\end{figure}

We plot   in Fig. \ref{fig-warp} the profile of the warp factors of Table \ref{tab-wf}, where one can realize that the differences between the GS and the HC models are present only  close to the origin. The $\sigma(r)$ factors both exponentially decrease, and $\lim_{r\to\infty}\beta(r)=1$. Moreover, the HC model preserves the regularity conditions at the origin $\sigma(0)=\partial_r\sqrt{\gamma(r)}\lvert_{r=0}=1$ and $\partial_r\sigma(r)\lvert_{r=0}=\beta(0)=0$ \cite{Silva:2012yj, Tyniakov}.

We study the vector gauge fields localization in the string-like models through the following action \cite{Oda1,Costa:2015dva, Diego, DM3, Coni2}:
\begin{eqnarray}\label{action-gauge-6D}
 S_{U(1)}=\int{\sqrt{-g} \, g^{MN}g^{RS}\mathcal{F}_{MN}\mathcal{F}_{RS}}d^6x,
\end{eqnarray}
where $\mathcal{F}_{MN}=\nabla_{M}\mathcal{A}_{N}-\nabla_{N}\mathcal{A}_M$, where the $\mathcal{A}_M$ denote gauge vector fields.
Hence the equation of motion reads \cite{Oda1,Costa:2015dva, Diego, DM3, Coni2}:
\begin{eqnarray}\label{eqm-gauge-6D}
 \frac{1}{\sqrt{-g}}\left(\partial_{S}\sqrt{-g} \, g^{SM}g^{RN}\mathcal{F}_{MN}\right)=0.
\end{eqnarray}
Next, adopting the following gauge conditions \cite{Oda1,Costa:2015dva, Diego, DM3, Coni2} $\partial_{\mu}\mathcal{A}^{\mu}=\mathcal{A}_{\theta}=0$ and $\mathcal{A}_{r}=\mathcal{A}_r(r, \theta)$, the Maxwell equations yield \cite{Oda1,Costa:2015dva, Diego, DM3, Coni2}:
\begin{subequations}\begin{eqnarray}\label{eqm-gauge2}
\left(\eta^{\mu \nu}\partial_{\mu}\partial_{\nu}+\frac{\sigma}{\gamma}\partial_{\theta}^2\right)\mathcal{A}_{r}&=0,\\
 \ 
\partial_{r}\left(\frac{\sigma^2}{\sqrt{\gamma}}\partial_{\theta}\mathcal{A}_{r}\right)&=0, \\
%\\ \nonumber
\left(\eta^{\mu \nu}\partial_{\mu}\partial_{\nu}+\frac{\sigma}{\gamma}\partial_{\theta}^2+\frac{1}{\sqrt{\gamma}}\partial_{r}\left(\sigma\sqrt{\gamma}\partial_{r}\right)\right)\mathcal{A}_{\lambda}&=0,
\end{eqnarray}
\end{subequations}
 with the usual Kaluza-Klein decompositions adopted  \cite{Oda1,Costa:2015dva, Diego, DM3, Coni2}
\begin{eqnarray}\label{kk-gauge}
 \mathcal{A}_{\mu}(x^{M},r,\theta)=\sum\limits_{n, l=0}^{\infty}a_{\mu}^{(n,l)}(x^\mu)\varphi_n(r)\ex^{il\theta}.
\end{eqnarray}
The wave equation for the gauge field on the brane for $l=0$ ($s$-wave) is given by \cite{Oda1,Costa:2015dva, Diego, DM3, Coni2} %and  $\mathcal{A}_{r}(x^{M},r,\theta)=\sum\limits_{l=0}^{\infty}\mathcal{A}_{\mu}^{(l)}(x^\mu)\varrho(r)\ex^{il\theta}$
 \begin{equation}
 \label{sl-spin1}
 \varphi_n''(r)+\left(\frac{3}{2}\frac{\sigma'}{\sigma}+\frac{1}{2}\frac{{\beta}'}{\beta}\right)\varphi_n'(r)+\frac{m_n^2}{\sigma}\varphi_n(r)=0,
 \end{equation}
where $\beta = \gamma / \sigma$. The solutions of Eq. \eqref{sl-spin1} are confined if it obeys the following equation:
 \begin{equation}
 \label{cond}
 \varphi_n'(0)=\varphi'(\infty)=0, \ \int_{0}^{\infty}{\sqrt{\sigma(r)\beta(r)}\varphi_n^*(r)\varphi_p(r) dr}=\delta_{np}. 
 \end{equation}
The gauge zero-mode is obtained by the substitution $m_n=0$ in Eq. \eqref{sl-spin1}, which has the normalized constant solution that obeys Eq. \eqref{cond}, reading 
 \begin{equation}
 \label{sl-0}
 \varphi_0(r)= \left(\int_{0}^{\infty}{\sqrt{\sigma(r)\beta(r)} dr}\right)^{-\frac{1}{2}}.
 \end{equation}
Hence, differently of the Randall--Sundrum (RS) \citep{RSI, bajc} model, the thin string-like model  has a confined gauge massless mode \cite{Oda1,Costa:2015dva, Diego, DM3, Coni2}, whereas the massive modes are not localized \cite{Oda1,Costa:2015dva, Diego, DM3, Coni2}.

 Let us now particularize the result to the GS model, where the results are analytically obtained. From the GS warp factor of the table \ref{tab-wf}, in the original radial equation, the GS warp factors turn Eq. \eqref{sl-spin1} into \cite{Costa:2015dva, Diego, DM3, Coni2}:
\begin{equation}
\varphi_n^{\prime\prime} - \frac{3}{2}c\varphi_n^{\prime} +m_n^2\ex^{cr}\varphi_n = 0.
\label{Sturm-Liouville-GS-gauge}
\end{equation}
The normalized massless mode reads to $\varphi_{0}(r)=\sqrt{\frac{c}{2\mathfrak{R}_0}}$, whereas  the massive modes have the form \cite{Costa:2015dva, Diego, DM3, Coni2}
\begin{equation}
\varphi_n(r) = \ex^{\frac{3}{4} cr}\left[B_{1_n} J_{3/2}\left(\frac{2m_n}{c}\ex^{\frac{1}{2}cr}\right) + B_{2_n} Y_{3/2}\left(\frac{2m_n}{c}\ex^{\frac{1}{2}cr}\right)\right],
\label{GS-MassiveModes-vector}
\end{equation}
where $B_{1_n}$ and $B_{2_n}$ are normalization constants to be fixed by the conditions \eqref{cond}.

In order to obtain a system where the derivative boundary conditions \eqref{cond} hold for the massive modes, we need to impose a cut off point
$r_{max}$ \cite{GS1, Costa:2015dva, Diego}. Hence, for the small mass regime $m_n < c$,  the following discrete spectrum $m_n=\frac{c\pi n}{2} \ex^{-\frac{c}{2} r_{max}}$ can be obtained, where $n>1$ is an integer \cite{Costa:2015dva, Diego}. %Costa:2013eua}. 
 
For the HC model, the results are only numerically  obtained \cite{Costa:2015dva}. However, due to the fact that asymptotically when $r\to\infty$ the HC model converts to the GS model, the eigenfunction differences are modified only at the origin \cite{GS1, Diego, Oda1, Costa:2015dva}. These differences are fundamental for the regularity conditions to hold, modifying the corrections to the Coulomb's law.

On the other hand, the transformation  of Eq.  \eqref{sl-spin1} into a Schr\"{o}dinger-like one is a very useful alternative approach. This is accomplished by a change of variables $dz =\int\sigma^{-\frac{1}{2}(r)} dr$ and $\varphi(z)=K(z)\varPhi(z)$ \cite{Costa:2015dva, Diego, DM3, Coni2}, with $K(z)$ being:
\begin{equation}
 \label{kz}
K(z)=\sigma^{-\frac{1}{2}}(z)\beta^{-\frac{1}{4}}(z).
 \end{equation} 

Thus, the Schr\"{o}dinger-like equation reads  \cite{Costa:2015dva, Diego, DM3, Coni2}:
\begin{equation}
 \label{sl-sch}
 -\ddot{\varPhi}_n(z)-U(z)\varPhi_n(z)=m^2_n\varPhi(z), 
 \end{equation} 
 where the dots denote derivatives with respect to $z$ coordinate, and the analogue quantum potential reads \cite{Costa:2015dva, Diego, DM3, Coni2}:
  \begin{equation}
 \label{sl-pot}
U(z)=\frac{1}{4}\left[2\frac{\ddot{\sigma}}{\sigma}-\left(\frac{\dot{\sigma}}{\sigma}\right)^{2}+\frac{\dot{\sigma}\dot{\beta}}{\sigma \beta}+\frac{\ddot{\beta}}{\beta}-\frac{3}{4}\left(\frac{\dot{\beta}}{\beta}\right)^2\right]
 \end{equation}
 
 In this case, the localization conditions in Eq.  \eqref{cond} are led to \cite{Costa:2015dva, Diego, DM3, Coni2}:
 \begin{equation}
 \label{condz}
 \frac{\dot{\varPhi}_n(0)}{\varPhi_n(0)}=-\frac{\dot{K}(0)}{K(0)}, \ \frac{\dot{\varPhi}_n(\infty)}{\varPhi_n(\infty)}=-\frac{\dot{K}(\infty)}{K(\infty)},  \ \int_{0}^{\infty}{\varPhi_n^*(z)\varPhi_p(z) dr}=\delta_{np}
 \end{equation}

Now, the localized gauge zero-mode into $z$ variable reads:
 \begin{equation}
 \label{sl-0z}
 \varPhi_0(z)= c_0 \sigma^{\frac{1}{2}}(z)\beta^{\frac{1}{4}}(z).
 \end{equation}
 being $c_0$ the normalization constant. Note that the $K(z)$ expression of Eq. \eqref{kz} is correlated with the gauge zero mode \eqref{sl-0z} by $K(z)=c_0/ \varPhi_0(z)$.

For the analytical GS model, Eq. \eqref{Sturm-Liouville-GS-gauge}, converted to the Schr\"{o}dinger-like equation \eqref{sl-sch},  has the form \cite{Costa:2015dva, Diego, Coni2}
 \begin{equation}
 \label{sl-sch-gs}
 -\ddot{\varPhi}_n(\tilde{z})+2 \tilde{z}^{-2}\varPhi_n(\tilde{z})=m^2_n\varPhi(\tilde{z}), 
 \end{equation}
where $\tilde{z}=\left(z+\frac{2}{c}\right)$. The normalized zero mode in the $\tilde{z}$ variable takes the form:
  \begin{equation}
 \label{sl-sch-gs-0}
\varPhi_0=\sqrt{\frac{2}{c}}\tilde{z}^{-1}.
 \end{equation}
The massive modes into $z$ variable are \cite{Costa:2015dva, Diego, Coni2}:
\begin{equation}
\varPhi_n(\tilde{z}) = \sqrt{\frac{2}{\pi}}\frac{\left(B_{3_n}-m_nB_{4_n}\left(\tilde{z}\right)\right)\sin\left(\tilde{z}\right)-\left(m_nB_{3_n}\left(\tilde{z}\right)+B_{4_n}\right)\cos\left(\tilde{z}\right)}{m\tilde{z}}
\label{GS-MassiveModes-vector-z}
\end{equation}
where $B_{3_n}$ and $B_{4_n}$ are normalized constant.

\begin{figure}[!htb] % Duas figuras lado a lado
\begin{minipage}[t]{0.45 \linewidth}
        \centering
%    \begin{subfigure}[b]{0.50\textwidth}
                \includegraphics[width=\linewidth]{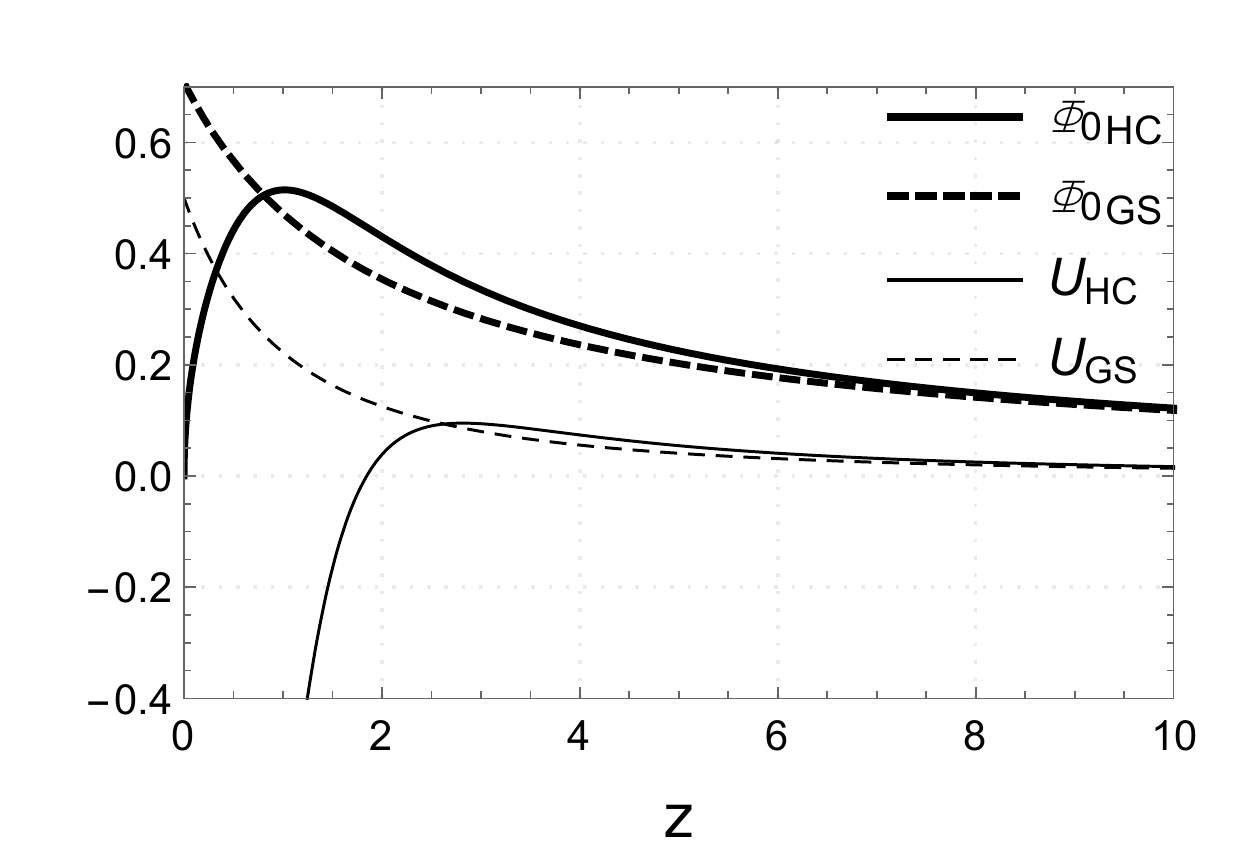}
                \caption{Massless mode $\varPhi_0(z)$ and the analogue quantum potential $U(z)$ for the GS and HC models. We set $c=1.0$.}
\label{fig-zero}
\end{minipage}
\quad
\begin{minipage}[t]{0.45 \linewidth}
%      \begin{subfigure}[b]{0.50\textwidth}
                \includegraphics[width=\linewidth]{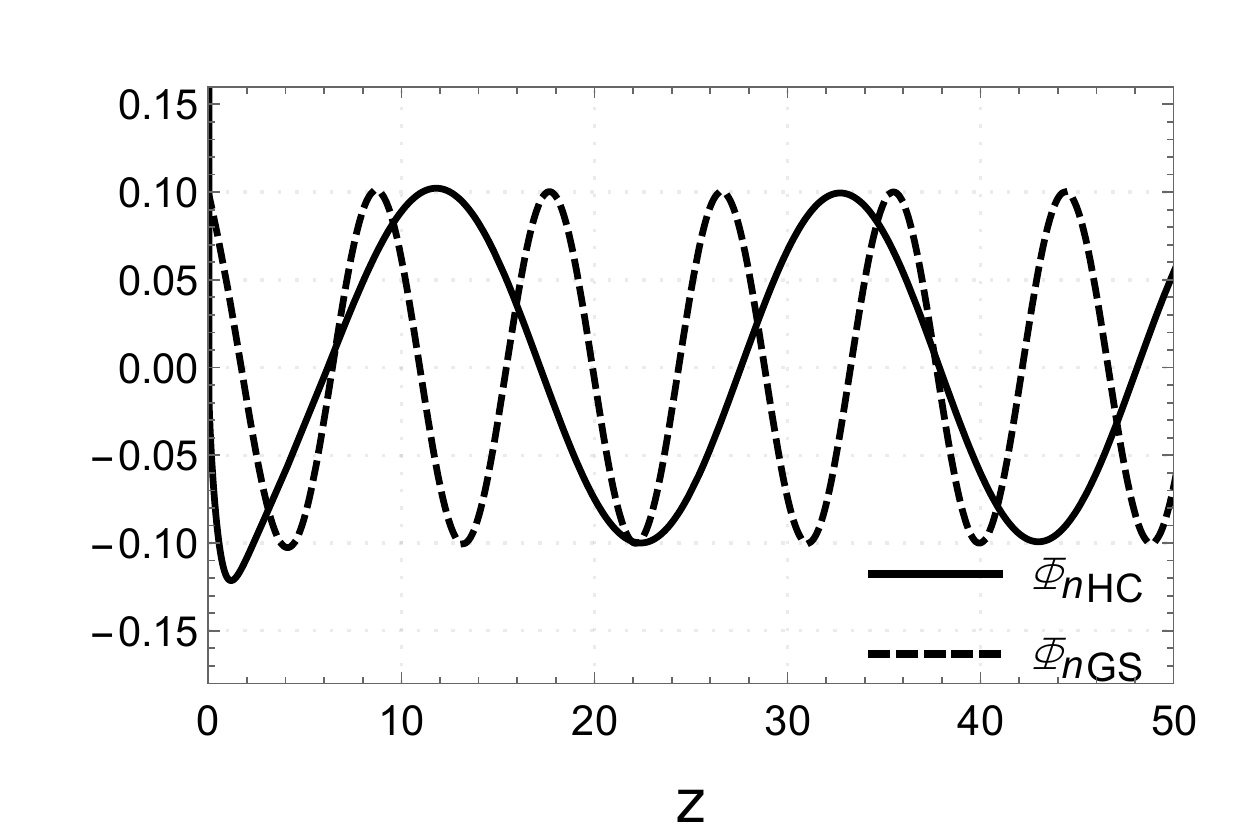}
                \caption{First excited massive mode $\varPhi_1(z)$  for $c=1.0$. For the GS model, $m_1= 0.7071$, and for the HC model, $m_1= 0.3083$.}
                \label{fig-mass}
\end{minipage}
\end{figure}

In order to illustrate these result, we plot in Fig. \ref{fig-zero} the massless modes $\varPhi_0(z)$ of Eq. \eqref{sl-0z} (thick lines) and the analogue quantum potential $U(z)$ of Eq. \eqref{sl-pot} (thin lines) for the GS (dashed lines) and the HC models (continuous lines). It is worth to emphasize  that both the zero modes are confined, being the GS centered at the origin, while the HC is displaced from it. The potential of the GS model is a barrel, whereas the potential of the HC model represents an infinite well. 

An example of massive mode is presented in Fig \ref{fig-mass}, where one notes that both oscillating function cannot obey the normalizing condition \eqref{condz}. Close to the origin, the massive modes of HC model displays a  non-periodical behavior, due to the brane displacement out from the origin \cite{Costa:2015dva, Diego, DM3, Coni2}.

%The values of the eigenfunctions  \eqref{GS-MassiveModes-vector-z} close to origin is a important result on order to obtain the correction to Coulomb's law. 

%\begin{equation}
%\varPhi_n(r) = \sqrt{\frac{2}{\pi}}\frac{\left[B_1-m_nB_2\left(z+\frac{2}{c}\right)\right]\sin\left(z+\frac{2}{c}\right)-\left[mB_1\left(z+\frac{2}{c}\right)+B_2\right]\cos\left(z+\frac{2}{c}\right)}{m \left(z+\frac{2}{c}\right)}
%\label{GS-MassiveModes-vector-z}
%\end{equation}

%%================================================================================================
%================================================================================================

\subsection{The corrections to the Coulomb's law on string-like models}

\begin{figure}[!htb] % Duas figuras lado a lado
%\begin{minipage}[t]{0.45 \linewidth}
        \centering
%    \begin{subfigure}[b]{0.50\textwidth}
                \includegraphics[width=.5\linewidth]{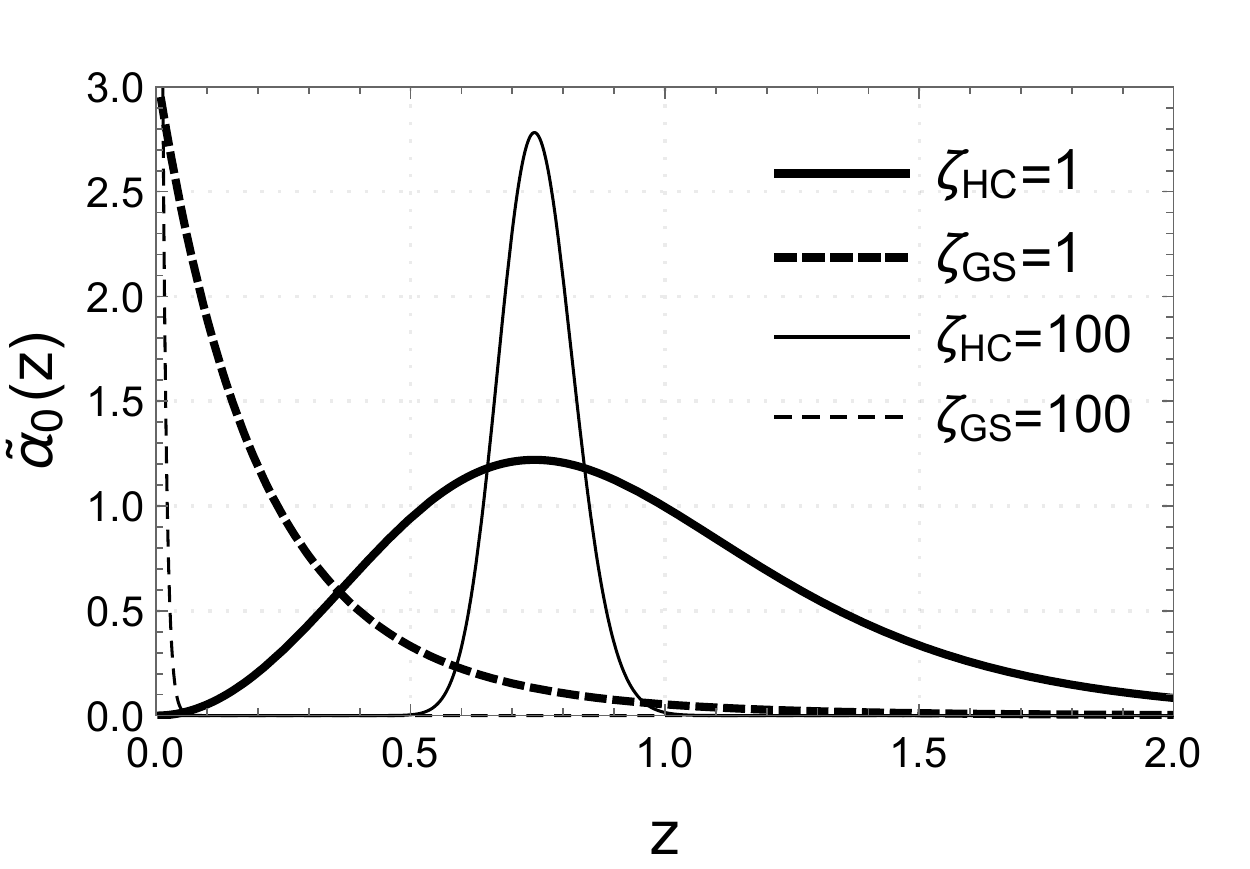}
                \caption{Left handed massless mode for the GS and HC models. The parameter $\zeta$ narrow the width of the fermion zero modes. We set $c=1.0$.}
\label{fig-alpha}
\end{figure}

The computation of the corrections for Coulomb's law in warped extra dimensional models was proposed in Ref. \cite{cou1,cou2,cou3}. In these references,  the interactions of the massive gauge modes with the massless left-handed fermion mode yield a correction of the Coulomb's potential $V_c(\x)$  in the form \cite{cou1, cou2, cou3}:
\begin{eqnarray}
V_c(\x)=\frac{\ex_0^2}{4\pi \x}\left[1+\Delta V_c(\x)\right] \ , \label{c5} %=\frac{\ex_0^2}{4\pi \x}+\int_{m_1}^{\infty}dm\frac{\ex^2_n}{4\pi \x}\ex^{-m \x}
\end{eqnarray}
where $\x$ and $\ex_0$ are  usual the norm of the position vector and the charge of the fermion trapped on the 4D brane, respectively. The corrections to the Coulomb's Law are given by the $\Delta V_c(\x)$ \cite{cou1, cou2, cou3}:
\begin{eqnarray}\label{deltav}
\Delta V_c(\x)=\int_{m_1}^{\infty}{dm}\ex^{-m \x}\left(\int_{-\infty}^{\infty}{dz}\tilde{\alpha}_{L_0}^{2}(z)\frac{\varPhi_n(z)}{\varPhi_0(z)}\right)^2 \ , %{c}^{-2}_{0}
\end{eqnarray}
where $m_1$ is the first excited gauge massive mode (given by the value of squared root of quantum analogue potential  maximum in Eq. \eqref{sl-sch}). These corrections are suppressed by the distance $\x$ and the mass $m$. The $c_0$ is the normalization constant of gauge zero mode,  obtained from Eq. \eqref{sl-0z}. Moreover, the $\tilde{\alpha}_{L_0}(z)$ is the fermionic normalized left zero mode in the string-like brane-worlds, which takes the form  \cite{DM3}:
\begin{equation}
\label{l0}
\tilde{\alpha}_{L_0}=C_{L_0}\sigma^{\frac{5}{4}\left(\zeta+3\right)}(z)\beta^{\frac{1}{4}\left(\zeta+3\right)}(z)
\end{equation}
where the $=C_{L_0}$ is the normalization constant for fermions and the parameter $\zeta>0$. The fermionic zero mode of Eq. \eqref{l0} is plotted in Fig. \ref{fig-alpha}, where we note that the parameter $\zeta$ narrows the width of the fermions zero modes.

 %Moreover. the term $\tilde{\alpha}_{L_0}^{2}(z) \varPhi_0^{-1}(z)=\left(C_{L_0}\right)^2 \sigma^{\frac{5}{2}\left(\zeta+\frac{14}{5}\right)}(z)\beta^{\frac{1}{2}\left(\zeta+\frac{5}{2}\right)}(z)$ can be obtained with the help of equations \eqref{l0} and \eqref{sl-0z}.

\begin{figure}[!htb] % Duas figuras lado a lado
%\begin{minipage}[t]{0.45 \linewidth}
        \centering
%    \begin{subfigure}[b]{0.50\textwidth}
                \includegraphics[width=.49\linewidth]{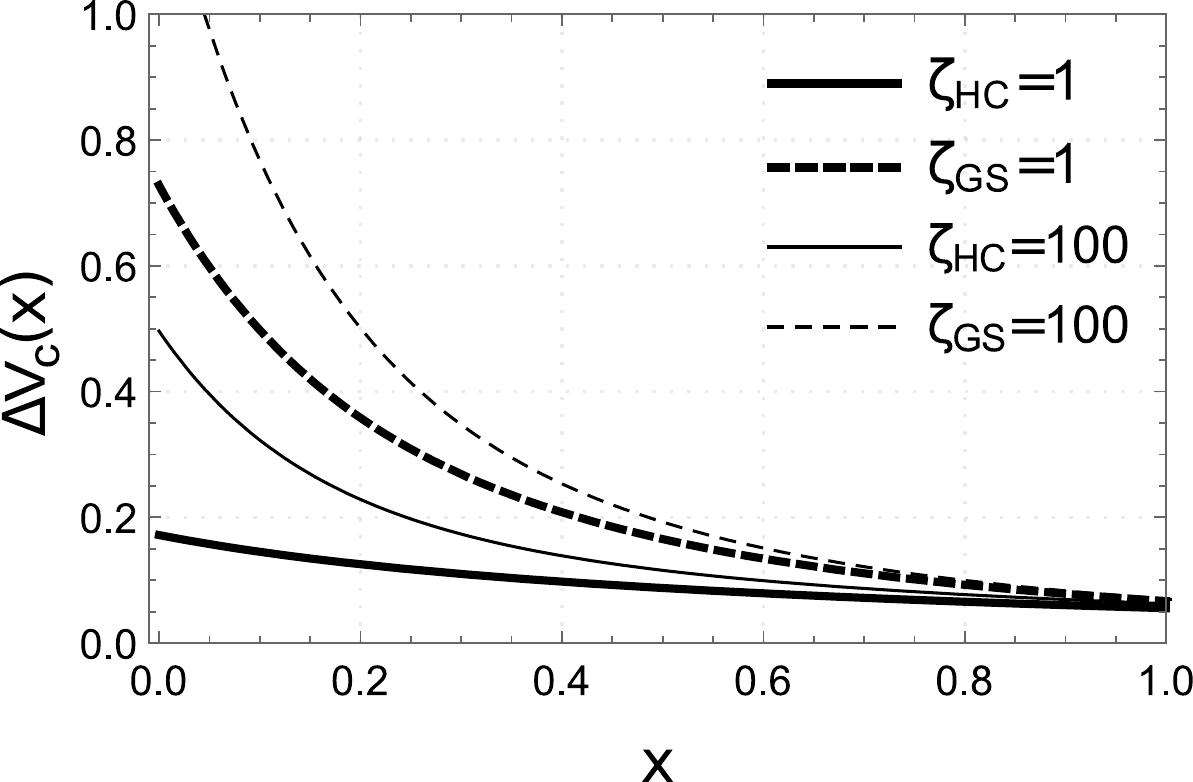}
                \includegraphics[width=.49\linewidth]{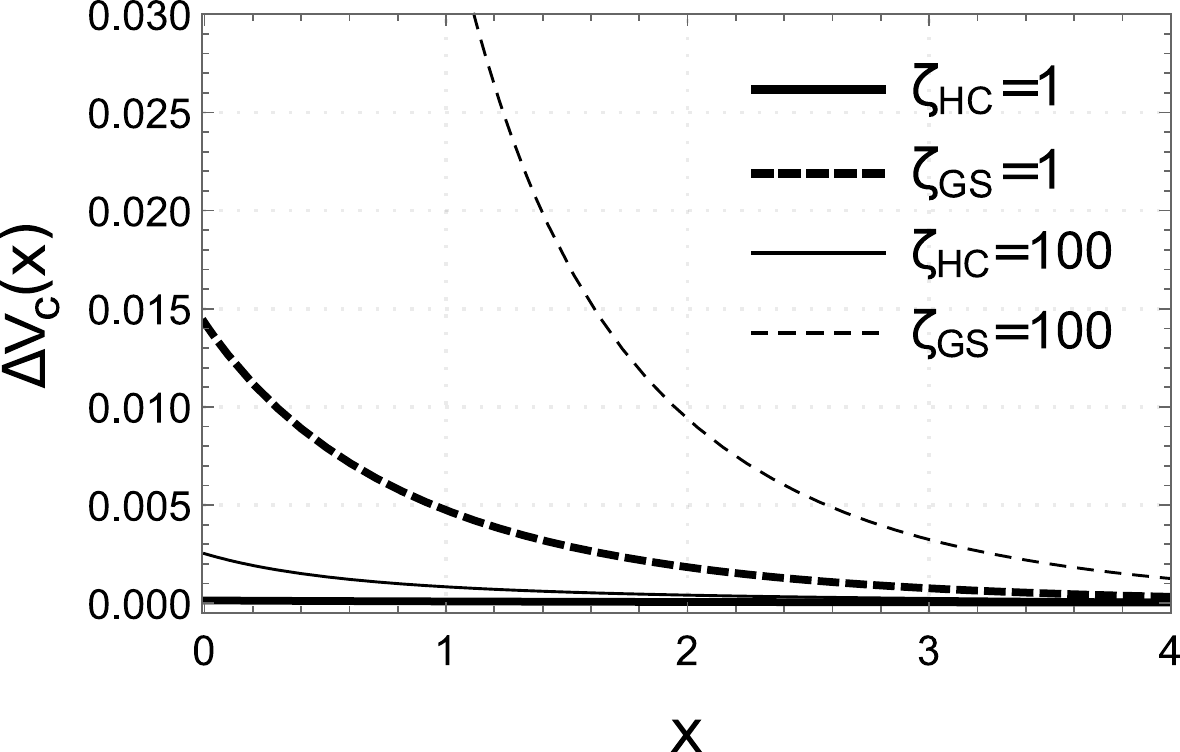}
                \caption{ $\Delta V_c(\x)$ correction to the Coulomb's Law. The left panel is set for $c=1.0$,  the GS has $m_1= 0.7071$ and the HC has  $m_1= 0.3083$. The right panel is set for  $c=0.5$, the GS has $m_1=0.3536$ and  the HC has $m_1=0.0318$.}
                \label{fig-delta}
%\end{minipage}
\end{figure}

Moreover, the plot of the Coulomb's potential corrections $\Delta V_c(\x)$ \eqref{deltav} is depicted in Fig. \ref{fig-delta}. For the left panel of Fig. \ref{fig-delta}, we set $c=1.0$ and the interval $z \in [0,100]$, whereas for the right panel of Fig. \ref{fig-delta}, we set $c=0.5$, being  the interval $z \in [0,200]$. Fig. \ref{fig-delta} shows that the corrections close to the origin for the GS modes are more expressive than in the  HC model. The parameters $\zeta$ and $c$ increase  the amplitude of the corrections.

%%================================================================================================
%================================================================================================

 In a qualitative way, Fig. \ref{fig-delta} shows a perspective for  these corrections, that can be adjusted to the experimental data. In addition, the corrections to Newton's law on regular string-like models were already studied in Ref. \cite{DM6}, where similar profiles were obtained.

%\newpage

\section{Monopoles in 6D warped thick brane-worlds}\label{sec-mono}
In this section we shall study how monopoles can physically constrain parameters that rule  the previously presented models. We start with the Georgi-Glashow model, consisting of an SU(2) gauge field $A_\mu$ and
a Higgs field $\phi$ in the adjoint representation, with the Lagrangian
\ba
\mathcal{L}=-\frac12 {\rm Tr}\,F_{\mu\nu}F^{\mu\nu}+{\rm Tr}[D_\mu,\phi][D^\mu,\phi]-m^2 {\rm Tr}\,\phi^2 - \lambda \left({\rm Tr}\,\phi^2\right)^2\,,
\ea
 where the covariant derivative reads $D_\mu=\partial_\mu + igA_\mu$ and   $ig F_{\mu\nu}=[D_\mu, D_\nu]$, for $A_\mu=A^a_\mu\sigma^a$ and $\phi=\phi^a\sigma^a$, for $\sigma^a$ being  generators of the SU(2) algebra. 

The effective Lagrangian reads
 \ba \hspace*{-0.2cm}S_{eff}&=&\int d\el{4}x \left[-\frac{1}{4}F_{\mu\nu}\el{a}F\el{a\mu\nu}+D_{\mu}\phi\el{a}D\el{\mu}\phi\el{a}-\frac{\lambda}{4}\Big(\phi\el{a}\phi\el{a}-\tilde{v}\el{2}\Big)\el{2}\right],\label{6}\ea  after a renormalization  $\phi\el{a}\mapsto \sigma^{-1/2}\phi\el{a}$, where the $\tilde{v}$ stands for the  spontaneous symmetry breaking dressed parameter, related to the  naked parameter by $\tilde{v}=ve\el{A}$. On the classical level, the model has two dimensionless parameters, given by the coupling constant, $g$, and $\lambda$. The scale is determined by $m^2$, whose negative values induces the SU(2) symmetry to be 
spontaneously broken into the Abelian U(1), by a non-zero vacuum expectation value of the Higgs field, given by  ${\rm Tr}\,\phi^2=\frac{v^2}{2}=\frac{m^2}{2\lambda}$. 
In the broken phase, the particle spectrum consists of a massless photon,
electrically charged $W_\pm$ bosons, and a neutral Higgs scalar, with respective masses 
\begin{eqnarray}
m_{W_\pm} &=& gv,\label{7}\\
m_H &=&\sqrt{2\lambda},\label{8}
\end{eqnarray}
  and massive magnetic monopoles.

The scalar and vector field for the monopole read \ba \phi\el{a} &=&\frac{{r}^{a}}{gr^2}H(gvr)\,\label{9}\\
 A_i &=&-\epsilon_{aij}\frac{r_j}{gr^2}\left[1-K(gvr)\right],\label{10}\ea where $H$ and $K$ are that describe a magnetic charge with localized energy.  The monopole mass reads $M_{m}=\frac{4\pi}{g\el{2}}h\left(m_H/m_{W_\pm}\right),$ where $h(0)=1$.  Eq. (\ref{7}) yields the warp factor  to be 
 realized by the monopole mass on the thick brane. The solution in (\ref{9}, \ref{10}) is physically a magnetic charge with localized energy, describing a particle with finite mass and a long-range magnetic Coulomb force between monopoles.

Determining the monopole radius  $R_{m}$ \cite{JP} is a task that 
makes the typical magnitude of $R_{m}$ to be chosen in order to balance the energy stored in the monopole  magnetic field, $g\el{2}/R_{m}$, and the energy provided by the monopole scalar field gradient  ($M_{W}\el{2}R_{m}$), yielding   $R_{m}\approx M_{W}\el{-1}$. On the other hand,  Eq. (\ref{7}) together with the functions $\tilde{v}$ yield \be  \tilde{R}_{m}\approx R_{m} \sigma^{-1/2}.\label{12} \ee  It is worth to realize that  the dressed monopole mass reads \be \tilde{M}_{m}=M_{m}\sigma^{1/2},\label{13} \ee   yielding the product  $\tilde{R}_{m}\tilde{M}_{m}=R_{m}M_{m}$ an invariant. 

Although the 't Hooft-Polyakov monopole solution is well known  obtained with a SU(2) initial gauge group  \cite{TP,poly}, whose gauge symmetry is  spontaneously broken at an extremely large mass scale,  of order $10\el{14}$ GeV \cite{WEI}. Therefore, the naked mass and the naked monopole radius read, respectively, $10\el{16}$ GeV and $10\el{-30}$ m, onto the brane core. Since the monopole radius depends on the  warp factor, the monopole radius increases out of the brane core. In order to analyse the 6D models of the previous section in this perspective, consider a cross section of the two extra dimensions with $\theta$ equals a constant, effectively providing the 5th dimension brane thickness. This can be justified by the expressions for the dressed parameters $\tilde{R}_m$ and $\tilde{M}_m$, since the warp factor in Eq. (\ref{6dmetric}) is what is taken into account for dressing such parameters. The previous analysis  of magnetic monopoles yield therefore  a constraint for the brane thickness.

Therefore, for the HC model, in order to avoid (unobserved) monopoles with mass scale of order TeV, taking into account that $R_{m}\sim 1/M_{m}$,  the condition $R_{m} \exp(-cr^*+\tanh(cr^*))\lesssim10\el{-15} {\rm cm}\label{22}$ must be imposed, where $r\el{*}$ denotes the brane `surface' along the extra dimension. This condition  reads 
\be 
\exp(-c\Delta_{\rm HC}/2+\tanh(c\Delta_{\rm HC}/2))\lesssim10\el{13},\label{23} 
\ee where $\Delta_{\rm HC}=2r\el{*}$ defines the brane thickness. Eq. (\ref{23}) implies $c\Delta_{\rm HC}\gtrsim 6.11588$ and
\be
\Delta_{\rm HC}\gtrsim 6.1158/c \label{plothc}
\ee
The parameter $c$ was identified in Ref. \cite{elko6DEPL} as an effective mass of dark matter particles $\sim 10^{-2}$ GeV, implying the brane thickness
\be
\Delta_{\rm HC}\gtrsim 6.1158\times 10^{-2}\;{\rm GeV}^{-1}=1.2037 \times 10^{-17} {\rm m}
\ee

The respective crosshatch region plot can be seen in Fig. 6, below.
 \begin{figure}[!htb] % Duas figuras lado a lado
%\begin{minipage}[t]{0.45 \linewidth}
        \centering
%    \begin{subfigure}[b]{0.50\textwidth}
                \includegraphics[width=.4\linewidth]{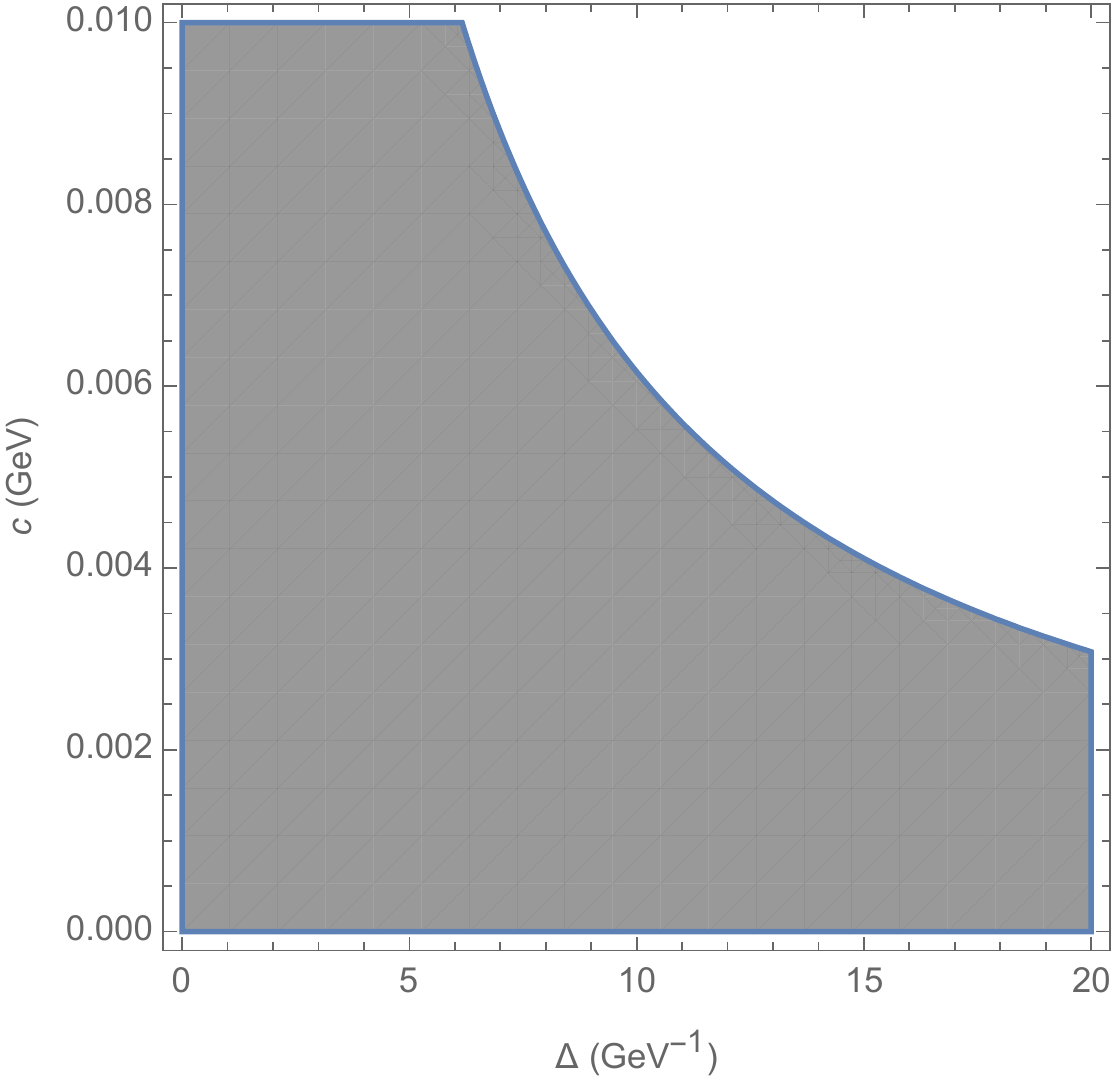}
                \caption{Region plot for the relation between the parameter $c$ in the HC model and 
                the brane thickness $\Delta$.}
\label{fig-alpha2}
\end{figure}
To determine the lower bound for the brane thickness, the parameter $c$ was identified in Ref. \cite{elko6DEPL} as an effective mass of dark matter particles $\sim 10^{-2}$ GeV, implying the brane thickness
\be
\Delta_{\rm HC}\gtrsim 6.4472\times 10^{-1}\;{\rm GeV}^{-1}=1.276 \times 10^{-16} {\rm m}. \label{thickhc}
\ee

\begin{figure}[!htb] % Duas figuras lado a lado
%\begin{minipage}[t]{0.45 \linewidth}
        \centering
%    \begin{subfigure}[b]{0.50\textwidth}
                \includegraphics[width=.4\linewidth]{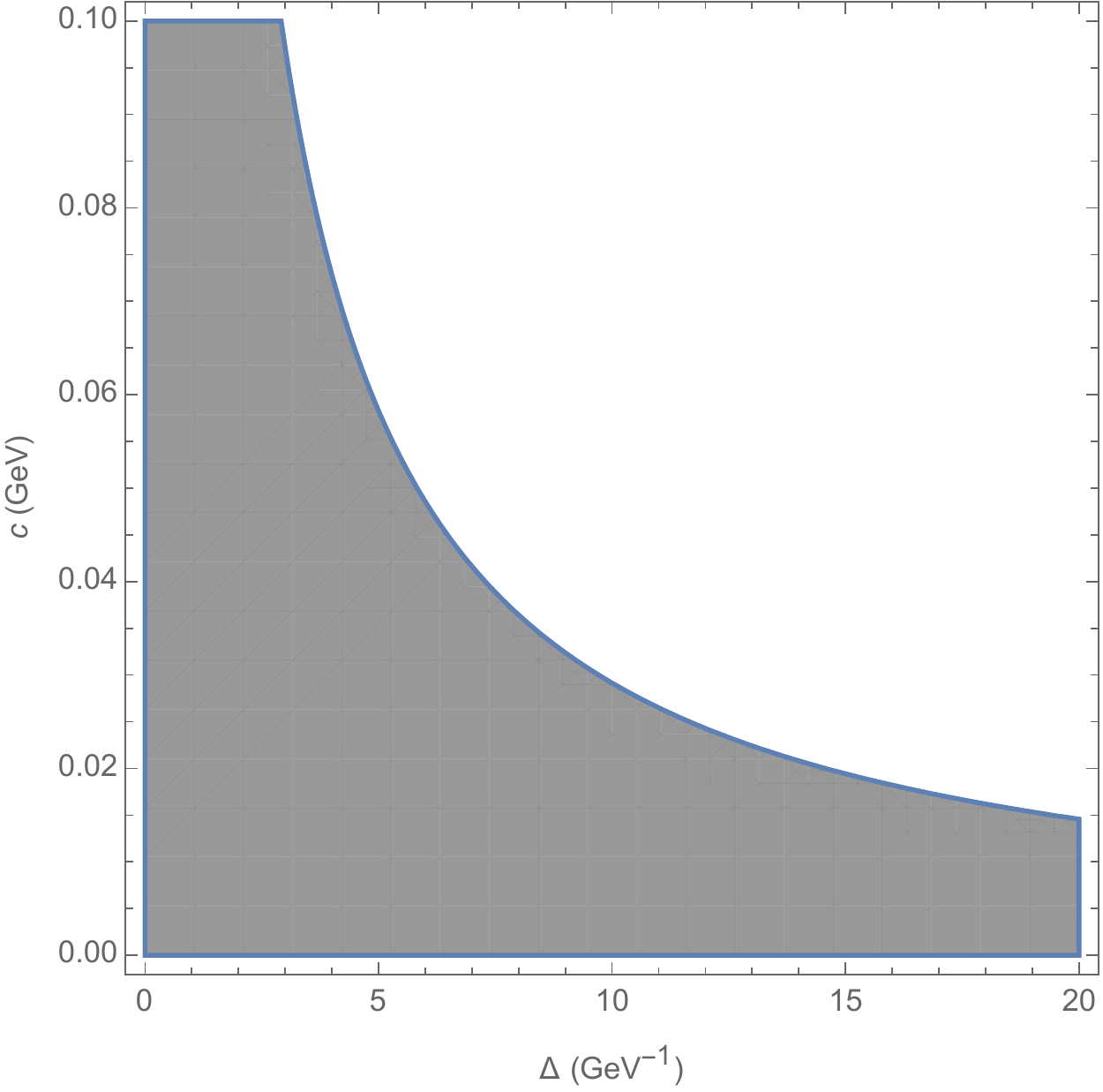}
                \caption{Region plot for the relation between the parameter $c$ in the GS model and 
                the brane thickness $\Delta$.}
\label{fig-alpha3}
\end{figure}

Analogously, we can also derive similar results for the GS model, wherein the condition $R_{m} \exp(-cr^*+\tanh(cr^*))\lesssim10\el{-15} {\rm cm}\label{22a}$   reads 
\be 
\exp(-c\Delta_{\rm GS}/2)\lesssim10\el{13},\label{2453} 
\ee yielding  
\be
\Delta_{\rm GS}\gtrsim 64.4724/c. \label{plotGS}
\ee
Again, as the parameter $c$ can be identified to the  effective mass of dark matter particles $\sim 10^{-2}$ GeV \cite{elko6DEPL}, it yields the brane width
\be
\Delta\gtrsim 29.9336\times 10^{-2}\;{\rm GeV}^{-1}=5.8914 \times 10^{-17} {\rm m}.\label{thickgs}
\ee
Similarly to the case of the HC model, the GS model has the region plot depicted  in Fig. \ref{fig-alpha3}.

%\newpage
One can realize that the brane width for the HC model, Eq. (\ref{thickhc}), and for the GS model, Eq. (\ref{thickgs}), are phenomenological bounds for 
the brane thickness into the 5th dimension, being respectively 3 and 2 orders of magnitude bigger than 
the theoretical bound of $2.0 \times 10^{-19}$ m.

\section{Conclusions}\label{sec-conclusions}

 After presenting string-like models in 6D, the gauge fields localization and the correction to the Coulomb's law, non observation of t'Hooft-Polyakov monopoles is employed to constrain the brane thickness for the HC and the GS model. 
The t'Hooft-Polyakov monopole radius and mass provided a phenomenological bound for the 6D brane thickness of the Gherghetta--Shaposhnikov  and the Hamilton Cigar  models. For the HC model, the derived width bound was $\Delta_{\rm HC}\gtrsim 1.276 \times 10^{-16} {\rm m}$, whereas for the GS model it yielded $\Delta_{\rm GS}\gtrsim 5.8914 \times 10^{-17} {\rm m}$. These results provide a tighter bound for the brane thickness into 5D. In fact, the lower bound was the 5D Planck scale,  $2.0 \times 10^{-19}$ m, which have been replaced by tighter phenomenological bounds. In this work, the corrections to the Coulomb's Law was only qualitatively studied, and its profile shares some similarities with corrections to the Newton's Law on string-like models. As a perspective, we want to adjust the scales for the $c$ parameter, in order to get more restrictive bounds based simultaneously in the corrections to the Coulomb's Law, the informational entropy, the monopole radius and mass, and many other phenomenological data.

\acknowledgments
RdR~is grateful to CNPq (Grant No. 303293/2015-2),
and to FAPESP (Grant No.~2017/18897-8), for partial financial support. DMD and CASA  is grateful to CNPq, to FUNCAP and to CAPES.

\end{document}